# High Energy Irradiation Effects on Silicon Photonic Passive Devices

Yue Zhou,[1,2] Dawei Bi,[1,2] Songlin Wang,[3,4] Longsheng Wu,[1,2] Yi Huang,[1,2] Enxia Zhang,[5,6] Daniel M. Fleetwood,[5] and Aimin Wu[1,2,7]

[1] State Key Laboratory of Functional Materials for Informatics, Shanghai Institute of Microsystem and Information Technology, Chinese Academy of Sciences, Shanghai 200050, China
[2] Center of Materials Science and Optoelectronics Engineering, University of Chinese Academy of Sciences, Beijing 100049, China
[3] Institute of High Energy Physics, Chinese Academy of Sciences, Beijing 100049, China
[4] Spallation Neutron Source, Science Center, Dongguan 523803, China
[5] Department of Electrical Engineering and Computer Science, Vanderbilt University, Nashville, TN 37235, USA
[6] enxia.zhang@vanderbilt.edu
[7] wuaimin@mail.sim.ac.cn

**Abstract:** In this work, the radiation responses of silicon photonic passive devices built in silicon-on-insulator (SOI) technology are investigated through high energy neutron and $^{60}$Co γ-ray irradiation. The wavelengths of both micro-ring resonators (MRRs) and Mach-Zehnder interferometers (MZIs) exhibit blue shifts after high-energy neutron irradiation to a fluence of $1 \times 10^{12}$ n/cm$^2$; the blue shift is smaller in MZI devices than in MRRs due to different waveguide widths. Devices with SiO$_2$ upper cladding layer show strong tolerance to irradiation. Neutron irradiation leads to slight changes in the crystal symmetry in the Si cores of the optical devices and accelerated oxidization for devices without SiO$_2$ cladding. A 2-μm top cladding of SiO$_2$ layer significantly improves the radiation tolerance of these passive photonic devices.

## 1. Introduction

Silicon photonics is an emerging technology that combines the advantages of low power consumption, high bandwidth, low-cost, small size, high integration, and compatibility with CMOS technology. Silicon photonic devices have great potential in optical communication, autonomous driving, biosensing, and microwave photonics. It is foreseeable that silicon photonics will become a fundamental technology of optical interconnection in high radiation environments such as satellites, spacecraft, space stations, nuclear reactors, and particle accelerators. When the devices are constantly exposed to a variety of high-energy radiation sources, their long-term reliability can be strongly affected. Therefore, the performance of these devices in radiation environments must be evaluated, and their reliability must be studied to ensure that the devices work well throughout mission life cycles.

It is well known that high-energy particles, such as neutrons and heavy ions, can cause significant displacement damage. The resulting defects can change optical properties of devices[1]. Neutrons are uncharged particles and will not cause Total Ionizing Dose (TID) damage directly[2]. $^{60}$Co γ-ray irradiation is another important source that is commonly used to study radiation effects on devices used in space applications. Gamma photons can produce secondary electrons by Compton scattering, creating displacement damage in the bulk material. In materials such as SOI, ionization damage and displacement damage can change optical properties[3].

Silicon photonic devices include various functional devices, such as silicon-based lasers, electro-optic modulators, photodetectors, filters, Wavelength Division Multiplexers (WDMs) devices, couplers, beam splitters, etc [4, 5]. In the last several decades, extensive researches show that radiation damage can degrade the performance of integrated circuits and electronic devices[6-8], III-V photodiodes[9, 10] and fibers[11-15]. However, only a few studies have investigated the impacts of radiation exposure on silicon photonic passive devices. Micro ring resonators (MRRs) and Mach-Zehnder interferometers (MZIs) are important passive devices that are extensively used as filters and WDMs. Silicon oxynitride MZIs and MRRs were exposed to 2.3 MeV α particles with a total fluence of $10^{15}$ cm$^{-2}$; a wavelength shift was observed due to an increase of $10^{-2}$ in the refractive index of SiON core and surrounding $SiO_2$ cladding[16]. The spectra of MRRs without passivation layers exhibited significant blue-shifts after exposure to X-rays or γ-rays, while variations of spectra were not observed for passivated MRRs[17, 18]. $^{60}$Co γ-ray irradiation also led to a slight wavelength shift of $SiO_2$-cladded SiC MRRs due to the radiation-induced refractive index changes in SiC core and $SiO_2$ cladding[19]. No wavelength shift was observed in a-Si MRR [20] because the refractive index of a-Si core and $SiO_2$ cladding was not changed after 15 Mrad(Si) of $^{60}$Co γ-ray exposure. V. Brasch et. al. showed that no significant radiation-induced degradation in SiN MRRs occurred after 18 to 99.7 MeV proton exposure with a fluence of $10^{10}$ cm$^{-2}$ [21]. Radiation effects in other photonic passive devices have also been evaluated, including lithium niobate waveguides[22], polymer waveguides[23], arrayed waveguide structures (AWS)[24], arrayed waveguide gratings (AWG)[25], InGaAsP/InP ring resonators[26] and Bragg gratings[27-29]. In these studies, wavelength shifts caused by radiation have been attributed to radiation-induced changes in the refractive index of the waveguide. Works have investigated radiation responses of bulk silicon[30] or silica[31] after irradiation. However, only a few studies have explored the mechanisms for these changes in separate waveguide core and cladding materials. Du et al. analyzed which constituent materials are responsible for the performance variation from the extrapolating results[19]. Analysis of the impacts of radiation exposure to integrated optical structures incorporated into waveguide devices is still lacking in some refractive index-sensitive applications.

Optical interconnection modules in the high luminosity Large Hadron Collider (HC-LHC), Spallation Neutron Source, or near a nuclear reactor, must function through high-intensity neutron and/or energetic particle irradiation. Grating couplers, MRRs, and MZI are typical passive devices that are used as integrated silicon photonics interconnection chips. In this work, we experimentally investigate the effects of neutron irradiation with a total fluence of $1\times10^{12}$ n/cm$^2$ and $^{60}$Co γ-ray irradiation with a total dose of up to 40 Mrad(Si) on silicon photonic passive devices, including grating couplers, MRRs, and MZIs. The mechanisms of radiation induced degradation in these devices are investigated through systematically pre- and post-irradiation characterization and data analysis of the devices.

## 2. Experimental details

The starting wafer includes a 220 nm top silicon layer and 2 μm buried silicon oxide (BOX). Grating couplers, MRRs with increased radii, and MZIs with increased imbalanced arms were fabricated in a CMOS platform, and then 2 μm $SiO_2$ cladding was formed via plasma-enhanced chemical vapor deposition (PECVD). Fig.1(a) shows a schematic view of the devices under radiation exposure. Fig. 1(b) is a 3D schematic diagram of devices with and without $SiO_2$ upper cladding. MRRs with different radii were fabricated on the same chip, including 3.1 μm, 4 μm, 6 μm, and 8 μm; the width of the waveguide core layer is 400 nm. The gap between the ring and bus waveguide is 100 nm. MZIs with different imbalanced arm lengths ΔL (from 75 μm to 125 μm) and waveguide width of 450 nm were fabricated to verify contributions of ΔL to radiation-induced responses. All devices are completely and uniformly exposed to radiation.

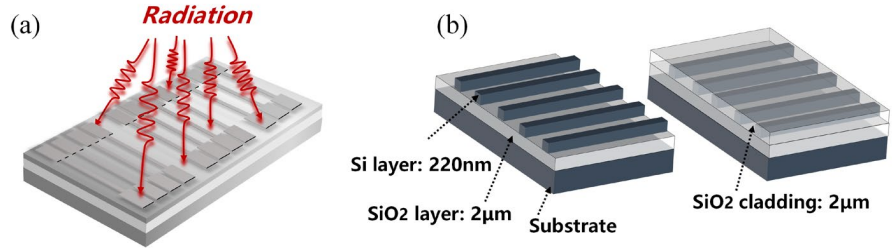

Fig.1. (a) Exposure of devices to high energy radiation. (b) 3D schematic diagrams of devices without cladding (left) and with SiO$_2$ cladding (right).

To evaluate the effects of high energy neutron irradiation on passive photonic devices, the devices were evenly divided into three groups. The first group was irradiated under a neutron beam with energies ranging from $5.8\times10^{-12}$ to $1.59\times10^{3}$ MeV to a total fluence of $10^{12}$ n/cm$^2$ with a flux of $10^8$ cm$^2$/s. The second group was exposed to a $^{60}$Co γ-ray source with energies of 1.17 MeV and 1.33 MeV at total doses of 23 Mrad(Si) and 40 Mrad(Si), respectively. The dose rate of the $^{60}$Co γ-ray source is 300 rad(Si)/s. The third group was used as a control. All the optical measurements in this work were performed at a temperature of $(21\pm0.2)$ °C. Significant wavelength shifts were observed in irradiated MRRs and MZIs.

To explore the effects of structural transformation mechanisms on optical property variation, irradiated samples were characterized by transmission electron microscopy (TEM), Raman spectra measurement, and X-ray Photoelectron Spectroscopy (XPS). Material characterization results indicate that high energy radiation can introduce displacement damage in core silicon and cladding SiO$_2$, which lead to the optical signal degradation.

### 3. Results

#### 3.1 Neutron irradiation on MRRs

MRR consists of a ring waveguide and a bus waveguide (inset of Fig. 2(a)). Interference occurs when the accumulated optical path difference in the ring is an integer multiple of wavelength ($\lambda$), and then the cavity is resonated. Even a tiny change of refractive index will lead to a peak shift.

Figs. 2(a) and (b) show measured normalized transmission spectra of MRR with the radius of 3.1 μm before (blue curve) and after (red curve) neutron irradiation to a fluence of $10^{12}$ n/cm$^2$. The Q-factors of both types MRRs are $\sim10^4$. Fig. 2(a) is the radiation response for MRR without cladding; the resonant wavelength of the irradiated MRR is blue shifted by 4 nm. For MRR with SiO$_2$ upper cladding in Fig. 2(b), a resonant wavelength blue shift of 615 pm is observed after irradiation. After irradiation, the Q-factors change by less than $\sim4\%$, which is within the measurement tolerance and standard deviation of device-to-device response. No significant radiation-induced degradation is observed in characterizations of loss, linewidth, or extinction ratios (ERs) for MRR devices under test in this work.

#### 3.2 Neutron irradiation on MZIs

The inset of Fig. 2(c) is the SEM image of an MZI. The incident light is equally divided into two separate arms, and constructive or destructive interference occurs when the accumulated phase difference is an integer multiple of $2\pi$ or an odd integer multiple of $\pi$, respectively.

To understand the effects of neutron irradiation on the MZIs, further experiments are carried out on devices without and with SiO$_2$ cladding. MZIs were exposed to neutrons with a total fluence of $10^{12}$ n/cm$^2$. Figs. 2(c) and (d) show the transmission spectra of MZIs with 95 μm imbalanced ΔL before (blue curve) and after (red curve) neutron irradiation. The peak wavelength blue shifts by 2.8 nm after neutron irradiation for MZIs without cladding in Fig.

2(c). In contrast, a wavelength shift of 370 pm is shown for MZIs with $SiO_2$ cladding (Fig. 2(d)). Meanwhile, loss, extinction ratios, and linewidth remain unchanged after radiation exposure.

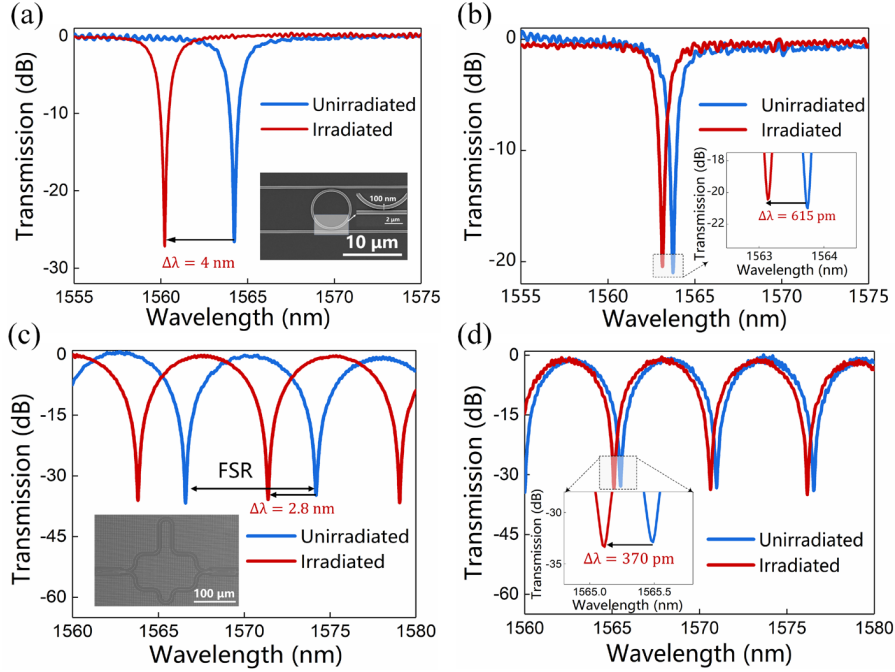

Fig. 2. Optical transmission spectra of MRRs (a) without cladding, (b) with $SiO_2$ cladding, before (blue curve) and after (red curve) $1\times10^{12}$ n/cm$^2$ neutron irradiation. Optical transmission spectra of MZIs (c) without cladding, (d) with $SiO_2$ cladding, before (blue curve) and after (red curve) $1\times10^{12}$ n/cm$^2$ neutron irradiation.

### *3.3 Neutron and Co-60 γ-ray irradiation on grating couplers*

Grating couplers are indispensable components for optical interconnect. It is essential to couple light from a single-mode fiber to a sub-micron waveguide. Fig. 3 shows the transmission spectra of the grating couplers without cladding (Fig. 3(a)) and with 2 μm $SiO_2$ cladding (Fig. 3(b)) under $10^{12}$ n/cm$^2$ neutron irradiation. The inset is the SEM image of the grating coupler. With or without upper cladding, the spectra of the grating couplers indicate that coupling loss and center wavelength are only slightly affected by irradiation.

To evaluate the radiation response of the devices and analyze the impact of different radiation sources, we also investigated the effects of $^{60}$Co γ-ray irradiation on grating couplers. Figs. 3(c) and 3(d) show their transmission spectra after 40 Mrad(Si) $^{60}$Co γ-ray irradiation. The influence on the performance of grating couplers is insignificant; neither a change in coupling loss nor a wavelength shift is observed. Hence, we can conclude that grating couplers are radiation-hardened to these radiations with high doses. Ahmed et. al. have shown that 1 MGy (100 Mrad) $^{60}$Co γ-ray irradiation hardly affects silicon MRRs[32]. Thus, the effects of $^{60}$Co γ-rays on MRRs and MZIs were not investigated in this work.

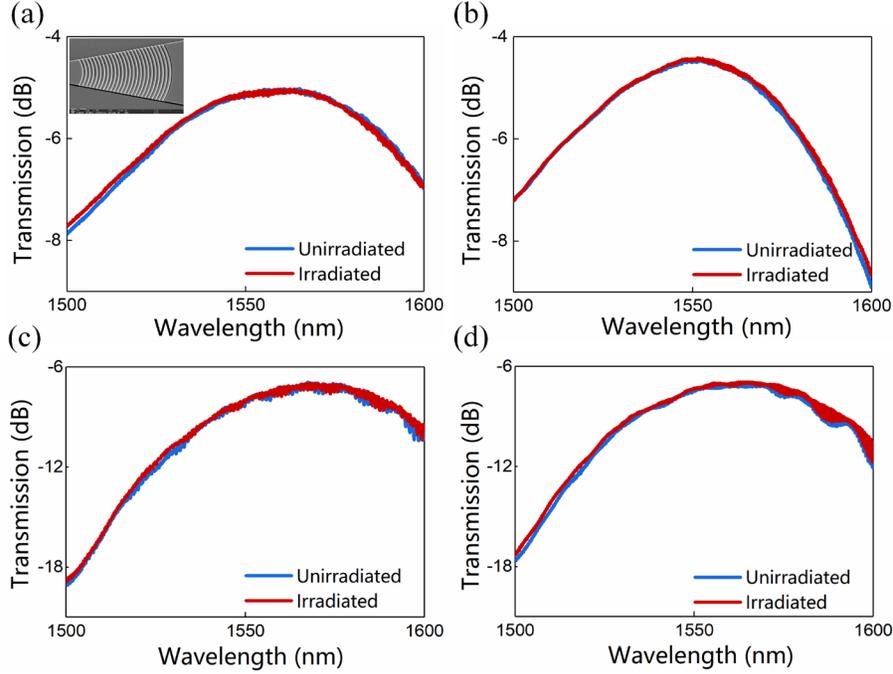

Fig. 3. Optical transmission spectra of grating couplers (a) without cladding and (b) with SiO$_2$ cladding before and after $1\times10^{12}$ n/cm$^2$ neutron irradiation, (c) without cladding and (d) with SiO$_2$ cladding before and after 40 Mrad(Si) γ-ray irradiation.

### 4. Discussion

Figs. 2-3 illustrate the radiation responses of the MRRs, MZIs and grating couplers. Neither linewidth nor wavelength of devices are changed significantly after irradiation, suggesting that grating couplers are robust to neutron and $^{60}$Co γ-ray irradiation. This may because γ-rays generate electron-hole pairs in Si/SiO$_2$ which will recombine with time at low dose rate and induce insignificant displacement damage in the material. In addition, the input light wave spectrum of grating coupler expands at the center wavelength according to the Fourier change due to limited length. On the other hand, they are not as sensitive to refractive index changes as MRRs and MZIs that are used as WDM, filters, etc. This can be responsible for the negligible changes in the transmission spectra within measurement resolution.

However, the blue shifts in peak wavelengths of MRRs and MZIs cannot be ignored in index-sensitive applications. To obtain further insight into the mechanisms responsible for the wavelength shifts, the average spectral shift $\Delta\lambda$ is used to infer the waveguide effective refractive index change via[17]:

$$\Delta n_{eff} = \frac{n_g \Delta\lambda}{\lambda_r} \qquad (1)$$

$$n_g = \frac{\lambda_r^2}{FSR \cdot \Delta L} \qquad (2)$$

So,

$$\Delta n_{eff} = \frac{\Delta\lambda \cdot \lambda_r}{FSR \cdot \Delta L} \qquad (3)$$

Here λ is the working wavelength and $\Delta n_{eff}$ is the change of effective refractive index of the waveguide mode. FSR is the period of the spectrum, named free spectral range. $\Delta L$ is the loop length for the MRR and the imbalanced arm length for MZI, respectively. As calculated by Eq. (3), the value of $\Delta n_{eff}$ is $1.15 \times 10^{-2}$ for MRR without cladding and $1.99 \times 10^{-3}$ for MRR with cladding. For MZI without cladding, $\Delta n_{eff}$ is $6.04 \times 10^{-3}$ and $1.13 \times 10^{-3}$ with cladding. In order to infer the refractive index changes of silicon core and $SiO_2$ cladding materials, different confinement factors are used to extract material index changes from optical measurements performed by Q. Du et al. via [19]:

$$\frac{\Delta n_{eff}}{\Gamma_{core} + \Gamma_{clad}} = (\Delta n_{core} - \Delta n_{clad}) \Gamma_{norm.core} + \Delta n_{clad} \quad (4)$$

$$\Gamma_{norm.core} = \frac{\Gamma_{core}}{\Gamma_{core} + \Gamma_{clad}}, \Gamma_{norm.clad} = \frac{\Gamma_{clad}}{\Gamma_{core} + \Gamma_{clad}} \quad (5)$$

Here, $\Delta n_{clad}$ and $\Delta n_{core}$ are cladding index change and core index change, respectively. $\Gamma_{core}$ and $\Gamma_{clad}$ are confinement factors of waveguide in the core and cladding regions. Fig. 4(a) shows the linear fits of measured data with $R^2$ of 0.99. From the extracting index, $\Delta n_{clad}$ and $\Delta n_{core}$ are $3.62 \times 10^{-3}$ and $-3 \times 10^{-4}$, respectively. Fig. 4(b) shows the simulated results in $\Delta$confinement factor of silicon core of MRRs with different radii by changing the calculated material index change through Finite Difference Eigenmode (FDE) solver in MODE Solution. The optical field of the waveguide becomes less localized due to the radiation induced material index changes and confinement ability degrades more severely with decreases in the bending radius.

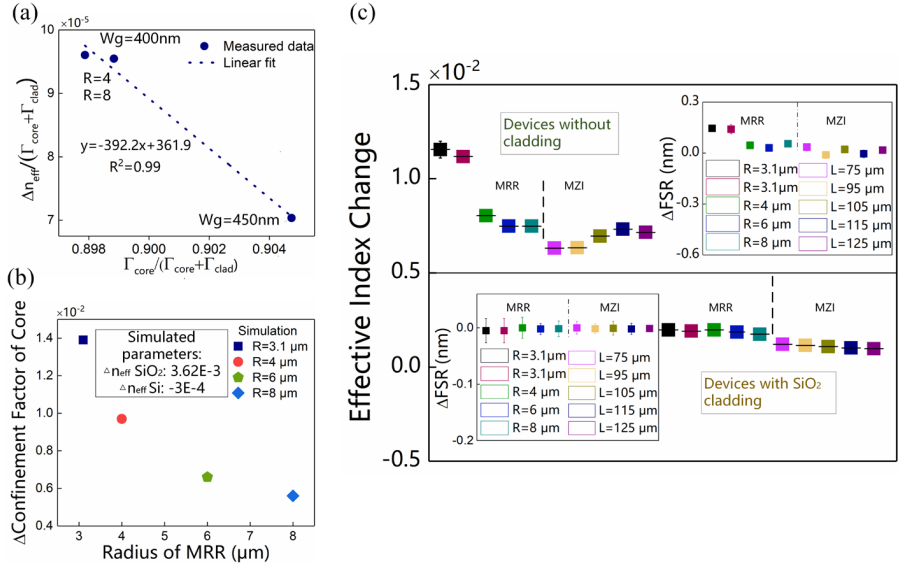

Fig. 4(a) Extracting material index changes from optical measurements. (b) Simulation of $\Delta$confinement factor of silicon core by MODE Solution. (c) Effective refractive index changes of devices without (up) and with $SiO_2$ cladding (down) before and after neutron irradiation, the insets are $\Delta$FSRs, respectively.

Fig. 4(c), which includes a 95% confidence interval, shows that the $\Delta n_{eff}$ of MRRs dependent on the different radii and the $\Delta n_{eff}$ of MZIs related to various arm lengths before and after neutron irradiation. The $\Delta n_{eff}$ values of the uncladded devices are greater than those of $SiO_2$ cladded devices. The $\Delta n_{eff}$ of MRRs increase with decreasing radii, which is consistent with simulated results in Fig. 4(b). MRR is more sensitive to high energy irradiation than MZI for both cladded and uncladded devices, which can be attributed to the smaller waveguide width of MRR, resulting in more sensitivity to material index changes. In addition, the effective refractive index changes of MZIs are affected slightly by arm lengths, which can be ignored. Insets in Fig. 4(c) show the values of $\Delta$FSR of devices with and without $SiO_2$ cladding. The

ΔFSRs of uncladded MRRs range from ~150 pm (R=3.1 μm) to ~25 pm (R=8 μm), and ΔFSRs of uncladded MZIs change much less (<20 pm). For devices with SiO$_2$ cladding, the ΔFSRs are less than ±10 pm. Thus, irradiation has a more severe impact on the uncladded devices.

After neutron irradiation, the increase of refractive index of cladding and the small decrease of core result in a reduction in effective refractive index and a blue shift in the wavelength. The observed wavelength shift after irradiation of the uncladded-device is more significant than that of the device with SiO$_2$ upper cladding. For uncladded devices, this trend may be attributed to the accelerated oxidation of unpassivated silicon under high energy irradiation[33, 34] and the displacement damage of core layer.

XPS and TEM were performed to analyze the surface oxidation on the uncladded devices. Figs. 5(a) and (b) exhibit detailed XPS spectra of Si 2p peak of uncladded materials. The black open circle denotes measured spectra, which are deconvolved into Si 2p$_{3/2}$ at 99.59 eV (blue curve), Si 2p$_{1/2}$ at 100.21 eV (green curve) and Si-O bond (purple curve). Before irradiation, quite a small fraction of Si-O bonds observed at 101.3 eV indicates that a small percentage of O atoms and Si atoms are combined into SiO$x$ (0<$x$<2)[35] due to the inevitable natural oxidation, and not into SiO$_2$ tetrahedral structure. After neutron irradiation, the fraction of Si-O bonds at 103 eV increases from 11% to 27%, indicating that the surface oxidation is accelerated by neutron irradiation. Fig. 5(c) is the TEM image of the irradiated-uncoated material. The thickness of the surface oxide layer is about 4 nm, resulting in lower silicon thickness. Consequently, the transmission spectra of MRRs and MZIs are blue shifted due to lower refractive index of silica (n$_{SiO_2}$=1.45) than that of crystal silicon (n$_{Si}$=3.45) [17].

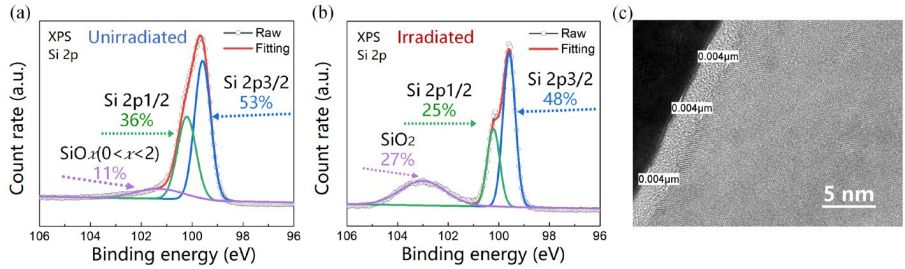

Fig. 5. XPS Si 2p core-level spectra of uncoated MRR device (a) before and (b) after 1×10$^{12}$ n/cm$^2$ neutron irradiation. (c) TEM image of the surface oxide of uncoated SOI after neutron irradiation.

Fig. 6 presents characterization results of the crystalline silicon in neutron-irradiated samples. In Figs. 6(a) and (b), two-dimensional electron diffraction patterns of the crystalline silicon via TEM compare the microstructure before and after neutron irradiation with a total fluence of 1×10$^{12}$ n/cm$^2$. The images show that the displacement damage has degraded the quality of electron diffraction patterns after irradiation. These changes result from loss of crystal symmetry of the silicon. Hence, neutron irradiation induced displacement damage in Si core is another main cause of the degradation in uncoated MRRs and MZIs.

The Raman spectrum of silicon is shown in Fig. 6(c), and the inset shows the partial enlarged view of the full width at half maximum (FWHM). After 10$^{12}$ n/cm$^2$ neutron irradiation, the Raman peak widens a little and shifts fractionally to a lower frequency, consistent with TEM results, suggesting that the crystalline symmetry is degraded in the Si core.

In general, when neutrons bombard the target material, the incident particles can displace atoms from lattice sites and create defects, leaving vacancies in the crystal. The decrease of atomic density and the surface accelerated oxidation result in a reduction of the refractive index [36, 37].

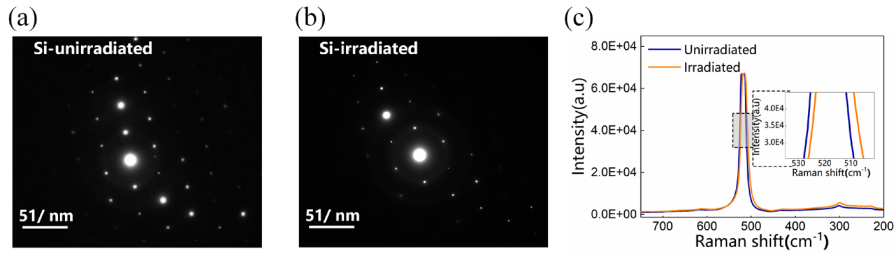

Fig. 6. TEM images of crystalline silicon core of uncoated device (a) before irradiation and after (b) $1\times10^{12}$ n/cm$^2$ neutron irradiation. (c) Raman spectrum of silicon is shown before and after $1\times10^{12}$ n/cm$^2$ neutron irradiation, the inset shows the partial enlarged view of the full width at half maximum.

To further explore the microscopic transformation of amorphous SiO$_2$ cladding, XPS measurements were performed. Figs. 7(a) and 7(b) show detailed XPS spectra of the Si 2p peak of amorphous SiO$_2$, which are deconvolved into two Si bonding states. The black open circle denotes raw data; the red curve is the sum of two convolved peaks. The pink curve corresponds to a Si-O-Si bridging bond at 103.5 eV, and the green curve is a Si-Si-O bond at 101.9 eV. As shown in Figs. 7(a) and 7(b), the fraction of Si-O-Si bridging bonds in amorphous SiO$_2$ decreases from 71.4% (before irradiation) to 68.1% (after irradiation), and the Si-Si-O bond percentage increases from 28.6% to 31.9%, consistent with the bond-breaking reactions[38]:

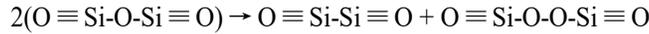

The breaking of the bridge bond corresponds to a reduction of the members in a ring. A-SiO$_2$ is a network of SiO$_4$ tetrahedra that includes n-membered planar rings, which denote *n* Si atoms and *n* O atoms in a ring[39]. When the multi-membered rings are subdivided into fewer-membered rings, the packing density of the material increases, and the broad distribution of the average Si-O-Si tetrahedral bridging-bond angle is reduced[38]. Fig. 7(c) shows a case that subdivision of a six-membered ring into two three-membered rings. Here the Si-O-Si angle $\varphi$ is given by F. L. Galeener[40]. After irradiation, the distributions of different types of rings in the internal structure are altered, which densifies the SiO$_2$ and increases the refractive index of amorphous cladding slightly.

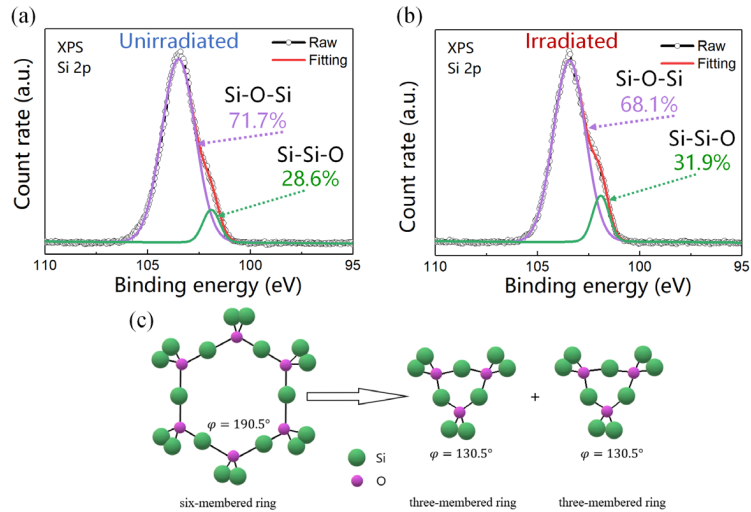

Fig. 7. XPS detailed spectra of Si 2p of SiO$_2$ (a) before and (b) after $1\times10^{12}$ n/cm$^2$ neutron irradiation. (c) Subdivision of a six-membered ring into two three-membered rings, green and purple circles denote silicon and oxygen atoms, respectively.

The Si/SiO$_2$ interface is another significant factor that affects the refractive index. More Si/SiO$_2$ interfaces are found in the integrated structure. A large concentration of dangling bonds exists at the interfaces. Near the interfaces, composition of SiO$_2$ species always vary, including Si$_2$O$_3$, SiO, and Si$_2$O[41]. Thus, the effective refractive index of the devices is also changed by irradiation.

## 5. Conclusions

We have performed a systematic investigation of high-energy and high fluence Co-60 gamma and neutron irradiation on the responses of silicon photonic passive devices, including grating couplers, MRRs, and MZIs. For MRRs and MZIs without SiO$_2$ cladding, the wavelength blue shifts more than the ones with SiO$_2$ upper cladding after neutron irradiation with a total fluence of $1\times10^{12}$ n/cm$^2$. The effective refractive index $\Delta n_{eff}$ of waveguide changes by about a ratio of $10^{-3}$, and the results show the waveguide size and bending radii dependence of these devices. We also find that devices with smaller waveguide width are more sensitive to high energy radiation, and MRRs with smaller bending radius experience larger index change. Material characterization results of TEM, Raman, and XPS illustrate that the combined effects of degradation in crystalline silicon due to displacement damage, surface accelerated oxidation, and the densification of the amorphous SiO$_2$ network change the refractive index.

We conclude that MRRs and MZIs are sensitive to variations of refractive index, making the devices susceptible to high-dose and high-energy irradiation. MZIs show higher radiation tolerance than MRRs due to greater waveguide width. Devices coated with upper cladding exhibit enhanced radiation resistance due to low possibility of radiation-enhanced oxidation in existing SiO$_2$ layer coating, indicating that a 2 μm SiO$_2$ upper cladding layer can significantly improves the radiation tolerance of these passive photonic devices.

**Funding.** Youth Innovation Promotion Association CAS (2021232); National Natural Science Foundation of China (61905269); Shanghai Sailing Program (19YF1456600), Shanghai Municipal Science and Technology Major Project (2017SHZDZX03).

**Acknowledgments.** The authors thank the support by Youth Innovation Promotion Association CAS, National Natural Science Foundation of China and Shanghai Sailing Program, Shanghai Municipal Science and Technology Major Project.

**Disclosures.** The authors declare no conflicts of interest.

**Data availability.** Data underlying the results presented in this paper are not publicly available at this time but may be obtained from the authors upon reasonable request.